\documentclass{acmart}

\AtBeginDocument{%
  \providecommand\BibTeX{{%
    \normalfont B\kern-0.5em{\scshape i\kern-0.25em b}\kern-0.8em\TeX}}}

\copyrightyear{2023}
\acmYear{2023}
\setcopyright{rightsretained}
\acmConference[Koli Calling '23]{23rd Koli Calling International Conference on
Computing Education Research}{November 13--18, 2023}{Koli, Finland}
\acmBooktitle{23rd Koli Calling International Conference on Computing
Education Research (Koli Calling '23), November 13--18, 2023, Koli,
Finland}\acmDOI{10.1145/3631802.3631854}
\acmISBN{979-8-4007-1653-9/23/11}

\acmPrice{15.00}
\acmISBN{978-1-4503-XXXX-X/18/06}

\begin{document}

\title{}

\author{Michael Guerzhoy}
\email{guerzhoy@cs.toronto.edu}
\affiliation{%
  \institution{University of Toronto}
  \city{Toronto}
  \state{Ontario}
  \country{Canada}
}

\title{``Medium-n studies" in computing education conferences}

\begin{abstract}
Good (Frequentist) statistical practice requires that statistical tests
be performed in order to determine if the phenomenon being observed
could plausibly occur by chance if the null hypothesis is false. Good practice also requires that a
test is not performed if the study is underpowered: if the number of observations is not sufficiently large to be able to reliably detect the effect one hypothesizes, even if the effect exists~\cite{simmons2011false}. Running underpowered studies runs the risk of false negative results.

This creates tension in the guidelines and expectations for computer science education conferences: while things are clear for studies with a large number of observations, researchers should in fact \text{not} compute p-values and perform statistical tests if the number of observations is too small~\cite{gelman2014beyond}. The issue is particularly live in CSed venues, since class sizes where those issues are salient are common

We outline the considerations for when to compute and when not to compute p-values in different settings encountered by computer science education researchers.

We survey the author and reviewer guidelines in different computer science education conferences (ICER, SIGCSE TS, ITiCSE, EAAI, CompEd, Koli Calling).

We present summary data and make several preliminary observations about reviewer guidelines: guidelines vary from conference to conference; guidelines allow for qualitative studies, and, in some cases, experience reports, but guidelines do not generally explicitly indicate that a paper should have at least one of (1) an appropriately-powered statistical analysis or (2) rich qualitative descriptions.

We present preliminary ideas for addressing the tension in the guidelines between small-n and large-n studies.

\end{abstract}

\maketitle

\section{Introduction}

In this paper, we study conference guidelines aimed at ensuring the replicability of published papers. Guidelines can enforce the use of statistics and language that encourage the authors to not draw conclusions not supported by the data.

A common desideratum for an empirical paper is that it be replicable. That is, if another researcher or educator tries the procedure described in the paper, they should get the same result. We can distinguish between strict or direct replication\footnote{\url{https://dictionary.apa.org/direct-replication}}, where the researcher or educators tries exactly the same experiments as the original authors, and conceptual replication\footnote{\url{https://dictionary.apa.org/replication}}, where the researcher tries to run experiments that are different from the original ones, perhaps because a setting is different, but using the same theoretical foundation and similar experimental methods. 

More expansive notions than replicability are \textit{generalizability} and \textit{external validity}~\cite{yarkoni2022generalizability}. A study is \textit{generalizable} if the theory its observations and/or experiments support can be applied in a novel setting. For example, a study (or experience report) showing active learning improves student outcomes is generalizable if one could follow the style suggested in the paper to improve student outcomes at different institutions, and possibly with different courses.

Recent large efforts to replicate well-known effects in Psychology have shown that fewer than two-thirds of well-known effects replicate~\cite{klein2018many}. 

General concerns with replicability and generalizability have become more salient in recent years~\cite{o2020ecologists}~\cite{ioannidis2005most} in social science. Some work has been done in CSed~\cite{ahadi2016replication}.

In this paper, we focus on the least expansive criterion: direct replicability. Whether and when ensuring direct replicability is desirable is a much less complex problem than addressing conceptual replicability, let alone generalizability.

Sanders et al.~\cite{ahadi2016replication} studied the use of statistical techniques in CSed (in contrast, we study the guidelines). Ahadi et al.~\cite{ahadi2016replication} explored CSed researcher attitudes and experiences to replication.

\section{Statistical Theory Summary}

\textbf{Small-n studies} are underpowered unless the expected effect is very large, which is uncommon in CSed. It is recommended that p-values not be computed for underpowered studies, since a significant result in an underpowered study can be misleading~\cite{gelman2014beyond}. A \textbf{Large-n} study, for large enough n, will generally be not underpowered (although issues of sampling bias and generalizability can remain). To guard against false-positive results obtained by ``eyeballing" the data, p-values for large-n studies \textbf{must be computed}. A p-value that is not computed when that was possible raises a red flag as to why it was not computed.

Studies where the number of samples is neither very small nor very large present a dilemma: it might be that a p-value must be computed, and it might be that they must not be computed.

Studies where p-values are not computed might be considered case studies in many guidelines.

\section{Guidelines in CSed Conferences}
We survey the guidelines for SIGCSE TS, ICER, EAAI, Koli Calling, ITiSCE, and CompEd. Details will be available in the poster. Some, but not all, conferences require justification for p-value computation. Many distinguish between case studies and ``CSed research", potentially implying that ``CSed research" is more replicable. Some explicitly emphasize ``statistical rigor'' in the research track. Koli Calling suggests, but doesn't require, the SIGSOFT Empirical Standards~\footnote{https://github.com/acmsigsoft/EmpiricalStandards}~\cite{ralph2021acm}.

\section{A way forward for medium-n studies}

We advocate for explicitly requiring at least one of (1) rich qualitative-like analysis or (2) appropriate statistical tests for all papers.

It will always be the case that some good papers do something between (1) and (2). However, usually, papers that don't contain rich qualitative analysis but do not have enough data to do quantitative analysis can be misleading.

\section{Conclusion}

In the culture in computer science education conferences, there is some ambiguity how to write papers with ``medium-sized" datasets. We do not offer definitive solutions. However, we believe authors, reviewers, and conference organizers should be deliberate about thinking about ``medium-n" papers.


\bibliographystyle{ACM-Reference-Format}
\bibliography{sample-base}

\end{document}